\documentclass{article}
\pdfoutput=1
\def\pd#1#2{ \frac{\partial #1}{\partial #2}}
\def\bvec{\left\{\begin{array}{c}} 
\def\evec{\end{array}\right\}}

\usepackage{graphicx}
\usepackage{cite}
\begin{document}

\title{A gas-kinetic scheme for the simulation of turbulent flows}


\author{Marcello Righi}

%

\maketitle
\begin{abstract}
Numerical schemes derived from gas-kinetic theory can be applied to simulations in the hydrodynamics limit, in laminar and also turbulent regimes. 
In the latter case, the underlying Boltzmann equation describes a distribution of eddies, in line with the concept of eddy viscosity developed by Lord Kelvin and Osborne Reynolds at the end of the nineteenth century. 
These schemes are physically more consistent than schemes derived from the Navier-Stokes equations, which invariably assume infinite collisions between gas particles (or interactions between eddies) in the calculation of advective fluxes. In fact, in continuum regime too, the local Knudsen number can exceed the value $0.001$ in shock layers, where 
%
%
gas-kinetic schemes outperform Navier-Stokes schemes, as is well known. 

Simulation of turbulent flows benefit from the application of gas-kinetic schemes, as the turbulent Knudsen number (the ratio between the eddies' mean free path and the mean flow scale) can locally reach values well in excess of $0.001$, not only in shock layers.
%
%
A further advantage of gas-kinetic schemes is that the fluxes are accurate to $\tau^2$, for instance in the scheme developed in \cite{xu2001gas} for the finite-volume discretization.
In laminar flow, this provides a better resolution of shocks and vortexes, whereas in turbulent flows, high-order fluxes allow for a better resolution of secondary flows in a manner comparable to higher-order turbulence models for the Navier-Stokes schemes.


This study has investigated a few cases of shock - boundary layer interaction comparing a gas-kinetic scheme and a Navier-Stokes one, both with a standard $k-\omega$ turbulence model.  
Whereas the results obtained from the Navier-Stokes scheme are affected by the limitations of eddy viscosity two-equation models, the gas-kinetic scheme has performed much better without making any further assumption on the turbulent structures.

\end{abstract}

\maketitle


\section{Introduction}

Numerical schemes based on the Navier-Stokes (NS) equations  have benefited for many years from the valuable work  of mathematicians and engineers. 
Aerodynamicists dispose of accurate and fast simulation tools, which can be applied to complex geometries and challenging flow conditions, providing physically consistent results in many cases.    

However, limitations still apply and appear difficult to overcome. They   concern the numerical model -- dependency of the results on numerical scheme and mesh, steeply increasing computational cost with the order of the scheme, reduction of accuracy at discontinuities -- or the physical model -- modelling the effect of unresolved turbulence on resolved flow 
in conditions far from local turbulent equilibrium, simulation of rarefied flow. The prediction of turbulent, hypersonic flow is a particularly challenging example.

Numerical schemes based on gas-kinetic theory, or Gas-Kinetic Schemes (GKS), might help in overcoming these limitations. Firstly, because the Boltzmann equation is a physically more accurate representation of fluid mechanics than the Navier-Stokes equations and secondly, because gas-kinetic can deal much better with the discontinuities that invariably at cells interface in most numerical approaches, since Godunov scheme (\cite{godunov1959difference}). 


Navier-stokes schemes split transport and collisions, i.e. advective and viscous fluxes. As such, advective fluxes are generated by the solution of the Riemann problem
which does not consider the effect of particle collision. The effect of collisions is added {\it a posteriori} by the viscous fluxes, calculated independently.
As long as the molecular relaxation time  is much smaller than the mean time scale, superposition of fluxes does not affect the physical consistence. However, wherever this assumption is not true,  as in the case of shock layers or of rarefied flows, the physical consistence of Navier-Stokes schemes becomes questionable. 
Despite the fact that Navier-Stokes schemes can predict shocks satisfactorily for many industrial applications, the prediction treats the shock merely as a discontinuity. Moreover, special treatment is often necessary to maintain the stability of the solution in presence of shocks - and this contributes to the dependence of the solution from the chosen method.


The modelling of turbulence can also benefit from the use of GKS. It is recognized since the publications of Lord Kelvin and Osborne Reynolds' works on turbulence, (refer to \cite{chen2003extended,chen2004expanded} and references therein) that the Boltzmann equation, also in the simpler form of the BGK model, can be used to describe not only the flow as distribution of particles but also the turbulent flow as distribution of eddies.  
Moreover, it is also recognized (\cite{chen2004expanded}) that the projection onto the physical space $(\bf x,t)$ of the BGK model generates a higher-order (in $\tau_t$) turbulent stress tensor. In particular a third order Chapman-Enskog expansion for $f$ generates a turbulent stress tensor of the second order, i.e. comparable to  non-linear turbulent stress models. These models are known to provide more accurate values  than linear models (\cite{pope2000turbulent,wallin2000explicit}). 

This study focuses on the GKS developed by Xu (\cite{xu2001gas}) which has also been investigated by other researchers (\cite{may2007improved,tang2011progress}) and has provided very good results in a number of cases, ranging from viscous-dominated, subsonic flows to hypersonics. It provides fluxes accurate to $\tau^2 $ and, consequently, a second-order turbulent stress tensor. 

This GKS has been implemented into an existing solver which uses a standard $k-\omega$ (\cite{wilcox2006}) turbulence model. The flow cases investigated are popular aerodynamic benchmarks - all in the continuum regime -  characterised by strong shock - boundary layer interaction - that is a flow condition where most two-equation turbulence models fail to predict shock position and extension of separated flow accurately.  
The equation for the turbulent quantities $k$ and $\omega$ are advanced in a segregate way. 



This paper presents a brief description of the GKS for laminar and  turbulent flow, followed by the numerical experiments and conclusions. 

\section{Gas-kinetic scheme }
\label{sec:gks}

A few gas-kinetic schemes for the solution of the Euler and the Navier-Stokes equations have been proposed in the 1990s (\cite{xu2001gas,may2007improved,mandal1994kinetic,chou1997kinetic,xu1994numerical} and references therein) as an alternative 
to the most popular schemes, which normally assume continuity of the flow or solve a  Riemann problem at cells interfaces. 


The main idea behind these schemes is to consider a discontinuous state across interfaces, re-construct the equilibrium and non-equilibrium distribution functions based on the macroscopic flow variables and calculate the evolution of the distribution functions during a time step $\Delta t$ integrating the BGK-Boltzmann equation. The macroscopic flow quantities are then recovered taking moments of the solution distribution function.



We use the macroscopic variables $\rho$,  $U = [u_1 \,\,\,u_2\,\,\, u_3]^T$, and $E$ to describe density, velocity and total energy of a gas. Instead of using the well-known Navier-Stokes equations, we write the BGK model following \cite{bhatnagar1954model}:  

\begin{equation}
\pd f t + u_i \pd f {x_i}  = {(g - f)\over\tau}
\label{3}
\end{equation}

\noindent where the summation convention holds,  $f$ is the gas distribution function, $g$ is the equilibrium state, a Maxwellian distribution, approached by $f$ and $\tau$ is the particles collision time, which is related to the molecular viscosity and heath coefficients of the gas. 
Although not explicitly indicated, it is assumed in \cite{xu2001gas} that the collision time can also include the effects of turbulence, beside those of molecular viscosity and numerical dissipation. 
The variable $\xi$ is related to the additional degrees of freedom of the gas molecules. $\xi$ has $K$ degrees of freedom, where: 

\begin{equation}
K=\frac{5-3\gamma}{\gamma-1}+1
\end{equation}

\noindent where $\gamma$ is the specific heat ratio. 
The equilibrium distribution is:

\begin{equation}
g = \rho \left( \frac{\lambda}{\pi} \right)^{\frac{K+2}{K}} e^{ -\lambda
\left(
(u_i-U_i)^2+\xi^2 \right)}
\end{equation}

\noindent where $\lambda = \frac{m}{2kT}$, $m$ is the molecular mass, $k$ is the Boltzmann constant, and $T$ is temperature.
The relation between macroscopic variables and gas distribution function is:  

\begin{equation}
\left\{
\begin{array}{c}
\rho \\
\rho U \\
\rho E \\
\end{array}
\right\}
=
\int\psi f d\Xi
\label{4} 
\end{equation}

\noindent where $\psi$ is:

\begin{equation}
\psi = 
\bvec
1 \\
U \\
{1\over 2} \left( {u_i}^2 +\xi^2 \right) 
\evec
\end{equation}

\noindent note that $d\Xi=du_1\,du_2\,du_3\,\xi^{K-1}\,d\xi$. 
Conservation of mass, momentum and energy during particle collision is expressed by: 

\begin{equation}
\int \left( g - f \right) \psi d\Xi = 0
\label{conservation}
\end{equation}

\noindent The BGK equation \ref{3} has an analytical solution:

\begin{equation}
f(x,y,z,t,u,v,w,\xi) = 
{1\over\tau} \int_o^t g(x',y',z',t,u,v,w,\xi) e^{-(t-t')/\tau}\,dt' + e^{-t/\tau} f_0 (x - ut,y - vt,z - w t) 
\label{eq:anasol}
\end{equation}
 
\noindent where $f_0$ is the initial gas distribution function, $x' = x - u(t-t'),\,\,\,y'=y-v(t-t'),\,\,\,z'=z-w(t-t')$.
The kernel of the GKS consists in expressing the distribution function $f$ at cells or volumes interfaces in order to assess the fluxes as functions of $f$. 
For instance the flux in direction $i$ at the interface between cells $n$ and $n+1$ can be expressed as a first moment of $f$:

\begin{equation}
F_i^{n+1/2}= \int_0^{\Delta_t} \int u_i^{n+1/2} \, \psi^{n+1/2} \, f(x^{n+1/2}) \, d\Xi\, dt
\label{eq:fluxesi}
\end{equation}

\noindent The distribution functions $f_0$ and $g$ in the \ref{eq:anasol} must be consistent with the macroscopic variables and their gradients. 
An important assumption in the derivation of the GKS is that whereas equilibrium distributions are Maxwellians, the non-equilibrium distribution are expressed as Taylor expansion of Maxwellian distribution. Assuming an interface normal to direction $1$ located at $x_1 = 0$, 
the initial equilibrium distribution is expressed as: 

\begin{equation}
g = \left\{\begin{array}{l}
g_0 \left(1+\bar a_i^l x_i-\bar A t \right),\,\,\,x_1<0 \\
g_0 \left(1+\bar a_i^r x_i-\bar A t \right),\,\,\,x_1>0 \end{array} \right. 
\label{expansiong0}
\end{equation}

\noindent where $g_0$ is a Maxwellian derived from a state $\left[ \rho_0 \,\,\, \rho {U_0}_i \,\,\, \rho_0 E_0 \right]$, which is an average state between left and right, obtained in a non-trivial averaging process, which fulfils the BGK model and the conservation laws (\cite{xu2001gas}). The initial distribution $f_0$ can be expressed as:

\begin{equation}
f_0= \left\{\begin{array}{l}
g^l (1+a_i^l x_i)-\tau \left( a_i^l u_i+A^l \right),\,\,\,x_1<0 \\
g^r (1+a_i^r x_i)-\tau \left( a_i^r u_i+A^r \right),\,\,\,x_1>0 \end{array} \right. 
\label{expansionf0}
\end{equation}

\noindent where $g^l$ and $g^r$ are Maxwellian distribution on both sides of the interface, which are indicated as {\it left} and {\it right}.
The choice of the terms used in the expansion \ref{expansionf0} is critical for the type of GKS. The terms proportional to $\tau$  represent the non-equilibrium parts in the Chapman-Enskog expansion (\cite{cercignani1988boltzmann}), whereas the expansion in the spatial directions $x_i$ is directly related to the formal accuracy of the resulting scheme. Moreover, one can have a directional splitting scheme by simply expanding in the direction normal to the interface, or a truly multi-dimensional scheme by considering the derivatives in all directions.

\noindent Each of the coefficients in the \ref{expansiong0} and \ref{expansionf0} is expanded as:

\begin{equation}
a_i = {a_i}_1 + {a_i}_2 u + {a_i}_3 v + {a_i}_4 w + {a_i}_5 (u^2+v^2+w^2+\xi^2)
\label{eqq:coeffexp}
\end{equation}

All the components of the coefficients are determined from compatibility relations with the macroscopic variables and the \ref{conservation}. The details can be found in \cite{xu2001gas}. The determination of all coefficients involves the solution of numerous (depending on the dimensions) linear systems and the evaluation of the {\it erfc} function, which contribute to the computational cost.
Inserting the \ref{expansiong0} and \ref{expansionf0} into the \ref{eq:anasol}, we obtain $f$:

\begin{eqnarray}
f  &=& 
       \left( 1 - e^{-t/\tau} \right) g_0 
  + \left( - \tau  + \tau  e^{-t/\tau}  +t\,e^{-t/\tau} \right) \left( h^l \, \bar{a_i}^l +  h^r \, \bar{a_i}^r \right) u_i  \, g_0 + \left( t - \tau  + \tau e^{-t/\tau}  \right) \bar{A} g_0 + \nonumber \\
   &+& e^{-t/\tau}  \left(  h^l g^l + h^r g^r - \left( t + \tau\right)  \left( u_i a_i^l h^l g^l + u_i a_i^r h^r g^r   \right) \right) - \tau  e^{-t/\tau}  \left( A^l h^l g^l + A^r h^r g^r \right)    
\label{resultingf}
\end{eqnarray}

\noindent where $h^l= H(U)$, $h^r = 1-H(u)$ and all coefficients are intended as series expansions in the form of  \ref{eqq:coeffexp}. 
Advective and viscous fluxes cannot be clearly separated in the \ref{resultingf}: this is a consequence of the fact that transport and collision of particles / eddies are considered simultaneously. Like in other gas-kinetic scheme the resulting fluxes appear as a series expansion in $\tau$, in this case accurate to the second order. 
Zero order terms provide the advective fluxes corresponding to the average state $g_0$, first order terms include the contribution to the viscous fluxes of the average state $g_0$ plus a correction to the advective fluxes, whereas second-order terms contain correction to the viscous fluxes and represent the real higher order contribution. 

\noindent The relaxation time $\tau$ is set as a function of the molecular viscosity of the gas plus an additional term proportional to the pressure jump across the interface. 

\begin{equation}
\tau = \frac{\mu}{p} + \frac{\left|p^r-p^l\right|}{\left|p^r+p^l\right|}\, \Delta t
\label{eq:tau}
\end{equation}

\noindent A known drawback of the BGK model is that it implies a unity Prandtl number; the heath flux must therefore be corrected for realistic gas / fluids (\cite{xu2001gas}).

\section{Turbulence modelling }
\label{sec:turbulence}

The present study is based on a simple modelling technique: the Reynolds approach, which resolves explicitly the mean flow and models the effects of all turbulent length scales - often referred to as RANS in its implementation with the Navier-Stokes equations.  
Turbulent quantities are modelled according to   $k-\omega$ model (\cite{wilcox2006}) two-equation models, which is a popular and accepted representative of this class of models.

\begin{equation}
\pd{\rho  K}t + \pd{\rho{u_j}  K}{x_j} = P 
- \beta^*\rho\omega  K
+ \pd{}{x_j} \left((\mu + \sigma^* \mu_t)\pd{ K}{x_j}\right)
\end{equation}

\begin{equation}
\pd{\rho\omega}t + \pd{\rho {u_j} \omega}{x_j} = 
 \gamma \rho {{\omega}\over { K}} P
+ \beta \rho {\omega}^2 
+ \pd{}{x_j} \left((\mu + \sigma \mu_t) \pd{\omega}{x_j}\right)
\end{equation}

\begin{equation}
\mu_t = \gamma^*\frac{\rho K}{\omega}.
\label{vistur}
\end{equation}

\noindent where $P$ is turbulence production term:

\begin{equation}
P = \tau_{ij} \pd{u_i}{x_j} 
\label{eq:pterm}
\end{equation} 

\noindent $\tau_{ij}$ is the turbulent stresses tensor: 

\begin{equation}
\tau_{ij} = \mu_t \left( 2 S_{ij} - \frac 23 \pd {u_k}{x_k} \delta_{ij} \right) - \frac 23 \rho k \delta_{ij} 
\label{eq:pterm}
\end{equation} 

\noindent and $S_{ij}$ is the strain rate: 

\begin{equation}
S_{ij} = \frac 12 \left( \pd{u_i}{x_j}+ \pd{u_j}{x_i} \right) 
\label{eq:pterm}
\end{equation} 

\noindent A typical choice for the parameters is:

\begin{equation}
\beta=\frac 3{40},\,\,\,\beta^*=\frac 9{100},\,\,\,\gamma=\frac 5 9,\,\,\,\gamma^*=1,\,\,\,
\sigma=\frac 1 2,\,\,\,\sigma^*=\frac 1 2
\end{equation}

%
%

\noindent 
%
The implementation of a turbulence model into a gas-kinetic scheme might seem straight-forward and practical steps taken to include the effects of turbulence into a GKS-based computation are really simple. 
Following \cite{xu2001gas} we can simply re-write the BGK model \ref{3} replacing the {\it molecular} relaxation time $\tau$ with a relaxation time $\tau^*$ which considers both molecular and turbulent phenomena.

\begin{equation}
\pd f t + u_i \pd f {x_i}  = {(g - f)\over\tau^*}
\label{eq:bgkturb}
\end{equation}

\noindent where

\begin{equation}
\tau^* = \tau + \tau_t
\label{tstar}
\end{equation}

\noindent where $\tau_t$ is the {\it turbulent} relaxation time, which can be expressed as a linear function of turbulent viscosity:  
 
\begin{equation}
\tau_t = \frac{\rho k }{\omega p} = \frac{\mu_t}{p}
\label{taut1}
\end{equation}

\noindent The \ref{3} can be rewritten as: 
 
\begin{equation}
\pd f t + u_i \pd f {x_i}  = {(g - f)\over\tau^*}
\label{gkst}
\end{equation}

\noindent In the practical calculations the \ref{eq:tau}, used in laminar flow, becomes:

\begin{equation}
\tau^* = \frac{\mu+\mu_t}{p} + \frac{\left|p^r-p^l\right|}{\left|p^r+p^l\right|}\, \Delta t
\label{eq:taut1}
\end{equation}

\noindent The turbulent relaxation time can also be expressed as a 
 non-linear function of the macroscopic variables as suggested by Chen (adapted from \cite{chen2003extended}):

\begin{equation}
\tau_t = \frac{{\rho\,k}/{\omega}}{p\, \left( 1 + \eta^2 \right)^{1/2}}
\label{taut2}
\end{equation}

\noindent where $\eta=S/\omega$ is the ratio between unresolved and resolved turbulence time scales, where $S$ is a scalar representing local velocity gradient and $\tau_0$ is the molecular relaxation time $\tau_0 = \mu / p$. 
The final expression for the  relaxation time including numerical dissipation is: 

\begin{equation}
\tau^* = \frac{\mu}{p}+\frac{{\rho\,k}/{\omega}}{p\, \left( 1 + \eta^2 \right)^{1/2}} + C \frac{\left|p^r-p^l\right|}{\left|p^r+p^l\right|}\, \Delta t
\label{eq:taut3}
\end{equation}

\noindent where the coefficient $C$ is determined heuristically (on average $C$ has been fixed at around $0.5$ for all turbulent simulations).

%
%
%
%
%
%

\section{Numerical experiments }
\label{sec:experiments}

All numerical experimental compare the results of a GKS and an Navier-Stokes scheme. Both schemes are implemented into a 2D structured, finite-volume spatial discretization.  The two schemes share the  reconstruction of the conservative variables at cells interfaces,  but differ in the evaluation of fluxes and in time stepping. 
Whereas the Navier-Stokes fluxes are obtained from Roe's approximate Riemann solver (advective) and from central differences (viscous), the GSK fluxes are obtained from Xu's scheme \cite{xu2001gas} extended to multi-dimensions.
Navier-Stokes are advanced by means of a third order RK whereas GKS uses a time-accurate single-step approach.
Both schemes use pre-conditioning (approximate LU-SGS based on the approximate factorization of the Navier-Stokes operator \cite{yoon1988lower} plus local time-stepping and multigrid acceleration (\cite{jameson1983solution}).  The approximate LU-SGS factorization had already been used with GKS operator in \cite{xu2005multidimensional}.
Most computations have been conducted with $CFL > 5$. 

The evaluation of GKS fluxes requires roughly three times longer than the Navier-Stokes. However, the evaluation of fluxes is required only once in a time step, whereas the Navier-Stokes requires multiple evaluations (depending on the scheme). Broadly speaking, the explicit schemes have very close time performances whereas the use of pre-conditioning makes the Navier-Stokes approximately twice faster. 

\subsection{RAE2822 and NACA 0012 airfoil in transonic flow }

Cases 9 and 10 of the measurements conducted by Cook (\cite{cook1979aerofoil}) on the RAE2822 supercritical airfoil as well as Harris' investigation  (\cite{harris1981two}) 
of the transonic flow around the NACA 0012 airfoil are arguably the most popular benchmarks for Navier-Stokes solvers and turbulence models, developed for transonic flow. In case 9 the boundary layer does not separate whereas in Case 10 and around the NACA 0012 at $M=0.800$, $Re = 9\times 10^6$ and $\alpha = 2.83^\circ$ the shock - boundary layer interaction leads to a large flow separation on the upper side of the airfoil. 

\begin{figure}[h!]
\includegraphics[width=78mm]{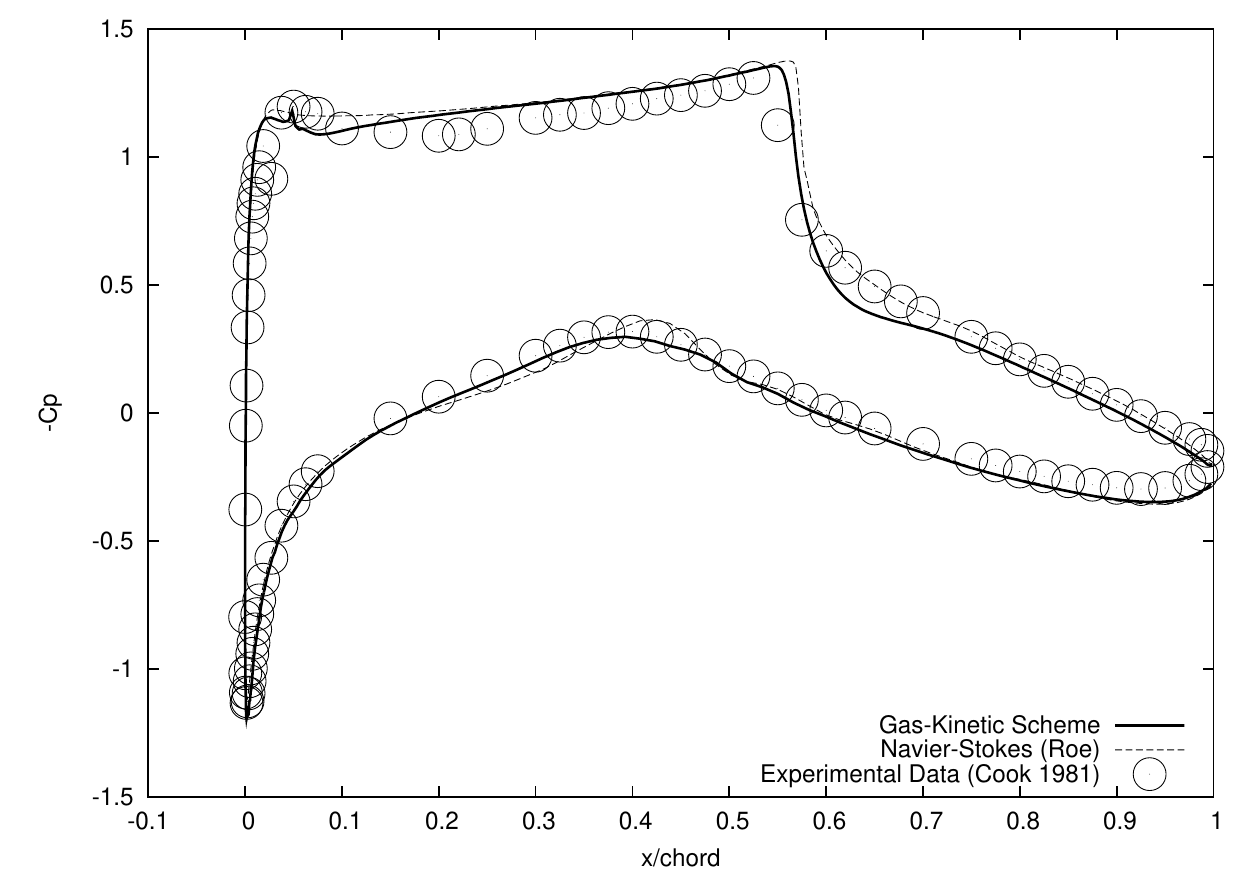}
\includegraphics[width=78mm]{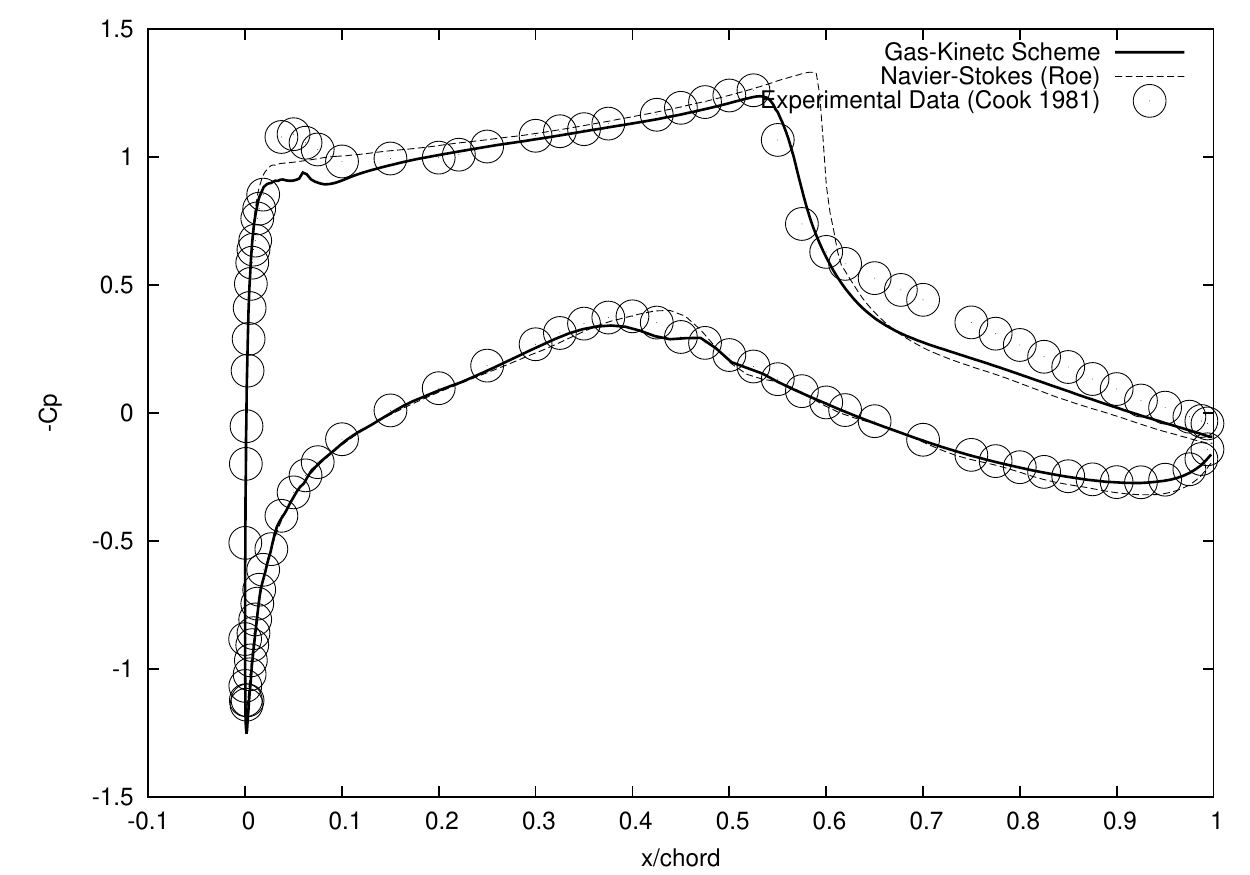}
\caption{Airfoil RAE2822, Case 9 (lhs) $Re = 6.2\times 10^6$, $M=0.725$,  angle of attack $\alpha =2.30^\circ$. Case 10 (rhs) $Re = 6.2\times 10^6$, $M=0.745$,  angle of attack $\alpha =2.30^\circ$. Pressure coefficient computed and measured. }
\label{fig:case910}
\end{figure}

\begin{figure}[h!]
\includegraphics[width=78mm]{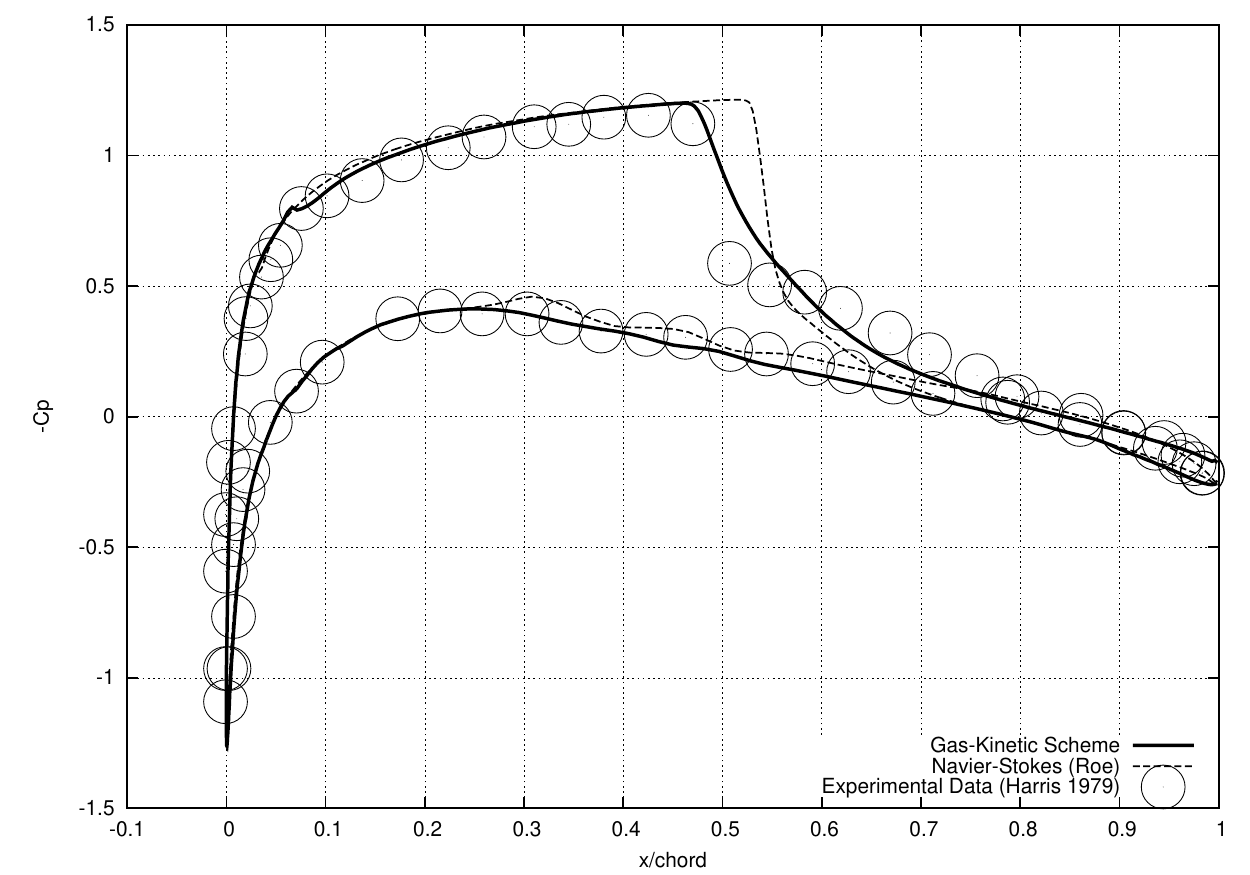}
\caption{Airfoil NACA 0012, $Re = 9.0\times 10^6$, $M=0.799$,  angle of attack $\alpha =2.26^\circ$. Pressure coefficient computed and measured. }
\label{fig:naca0012m08}
\end{figure}

It is well known that two-equation turbulence models, such as the $k-\omega$, become less and less accurate as the separated region grows. In the two separated cases shown here, the size of the separated area is typically underestimated and the position of the shock predicted slightly downstream (refer for instance to \cite{wallin2000explicit}). 
Figures \ref{fig:case910}  and \ref{fig:naca0012m08} show the pressure distribution obtained with the Navier-Stokes and the GKS schemes. The computational meshes are C-type with $625 \times 125$ cells (RAE 2822) and $624 \times 128$ cells (NACA 0012),  with $y_1^+ < 1$ (resolution of the boundary layer in wall units) in both cases.
Results of Navier-Stokes and GKS are comparable for Case 9 (figure \ref{fig:case910} lhs), whereas the GKS shows a higher accuracy in the capture of the shock - boundary layer interaction in Case 10 and in the NACA 0012 case (figure \ref{fig:case910} rhs and figure \ref{fig:naca0012m08} respectively), in that the flow separation is predicted much more accurately. 
The poor prediction of separated flows by the $k-\omega$ model 
can be improved by the replacing the linear expression for the  eddy viscosity with algebraic relations for the components of the Reynolds stresses tensor (\cite{wallin2000explicit}).

\subsection{Airfoil NACA 64A010 in transonic flow at high angle of attack}

\begin{figure}[h!]
\includegraphics[width=78mm]{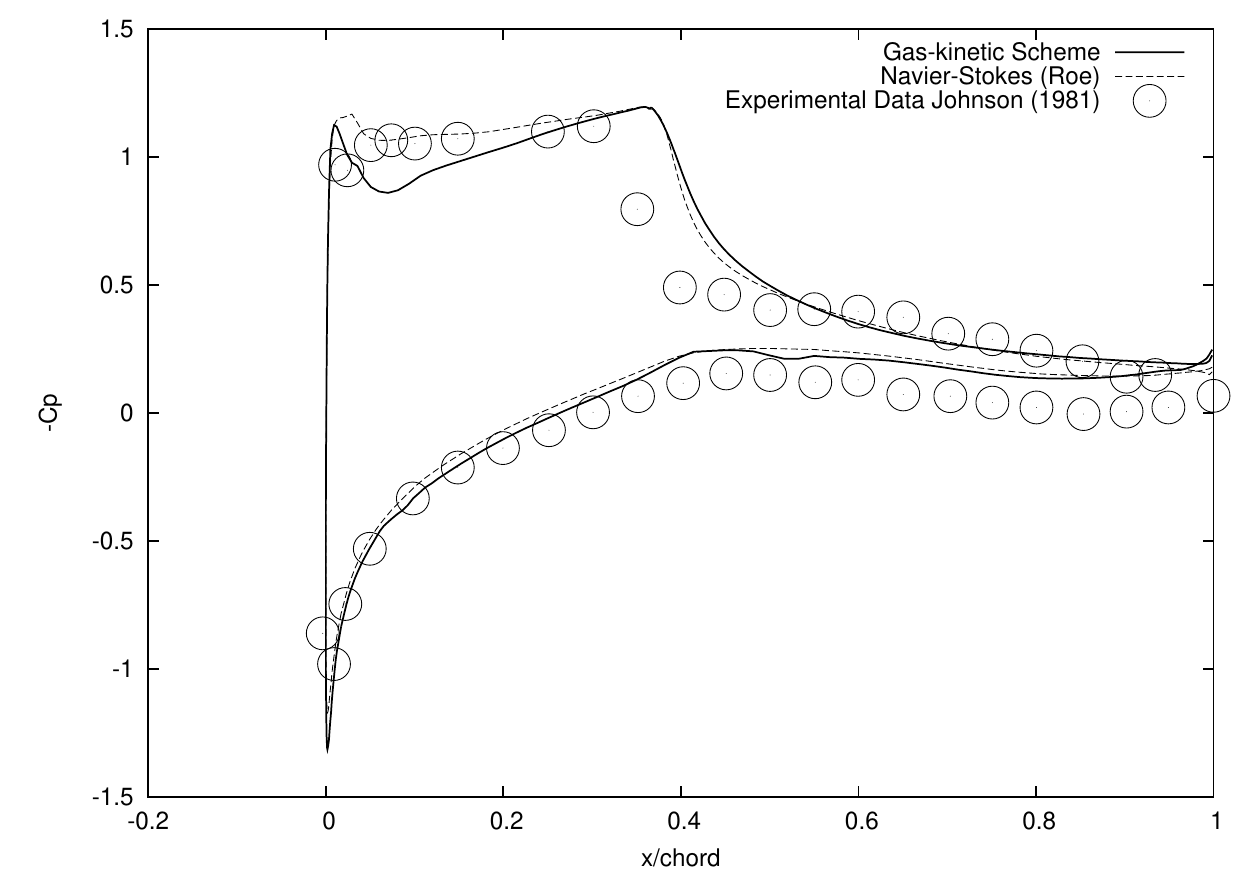}
\caption{Airfoil NACA 64A010, $Re = 2.0\times 10^6$, $M=0.75$,  angle of attack $\alpha =6.2^\circ$. Pressure coefficient computed and measured. }
\label{fig:naca64a010}
\end{figure}

\begin{figure}[h!]
\includegraphics[width=108mm]{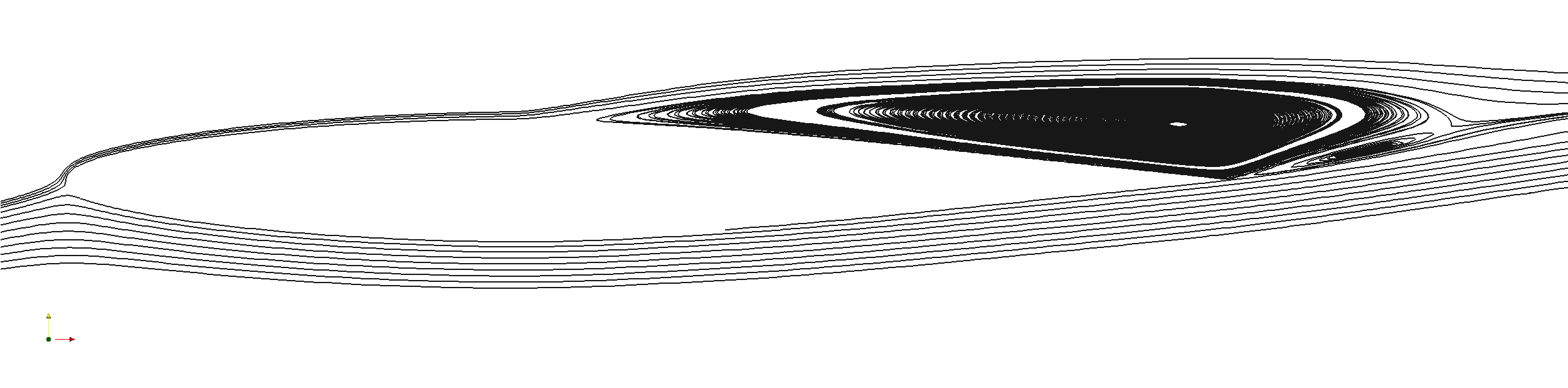}
\caption{Airfoil NACA 64A010, $Re = 2.0\times 10^6$,  $M=0.75$, angle of attack $\alpha = 6.2^\circ$. Streamlines showing the large separation induced by the shock obtained from the Navier-Stokes solutions. }
\label{fig:64a010ns}
\end{figure}

\begin{figure}[h!]
\includegraphics[width=108mm]{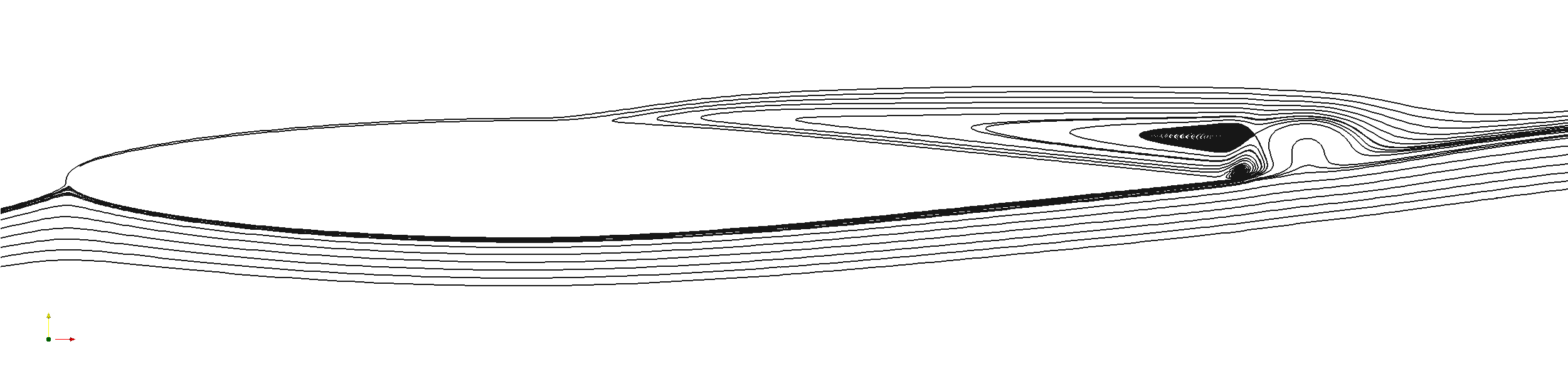}
\caption{Airfoil NACA 64A010, $Re = 2.0\times 10^6$,  $M=0.75$, angle of attack $\alpha = 6.2^\circ$. Streamlines showing the large separation induced by the shock obtained from the the Navier-Stokes (lhs) and GKS (rhs) solutions. Details of streamlines around the trailing edge. }
\label{fig:64a010gks}
\end{figure}

\begin{figure}[h!]
\includegraphics[width=78mm]{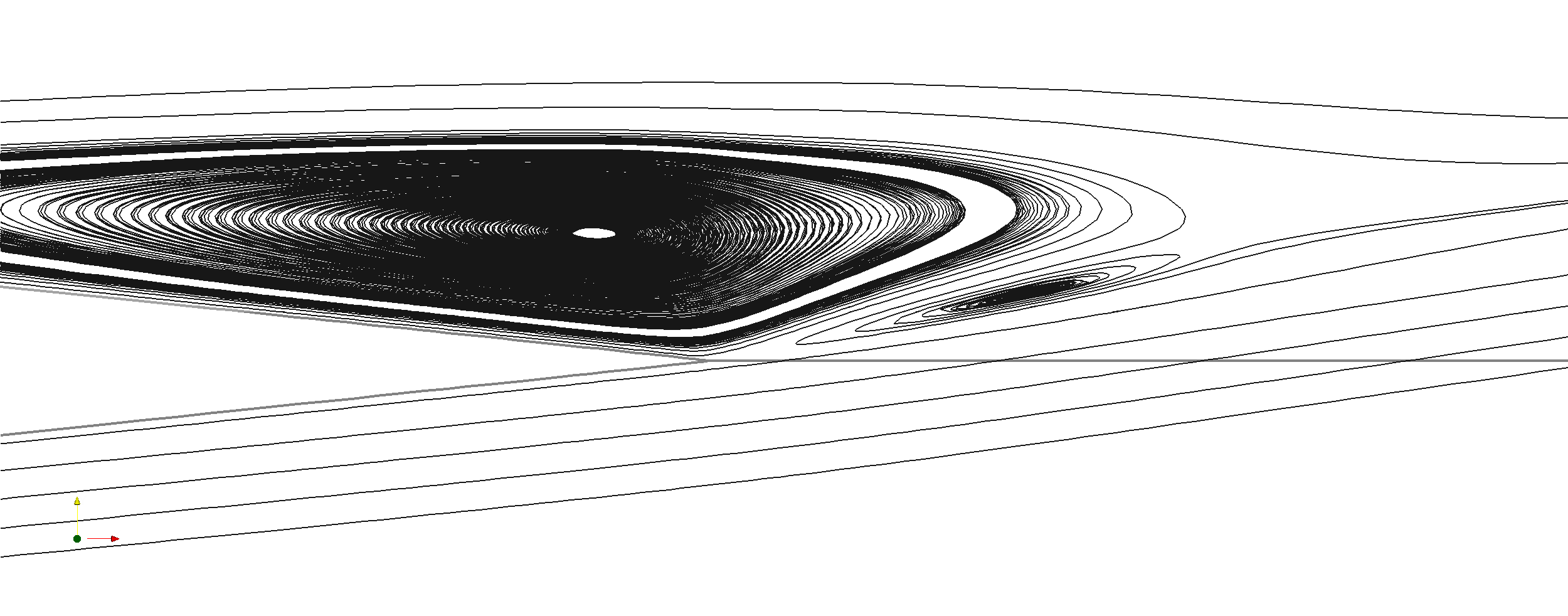}
\vspace*{2mm}
\includegraphics[width=78mm]{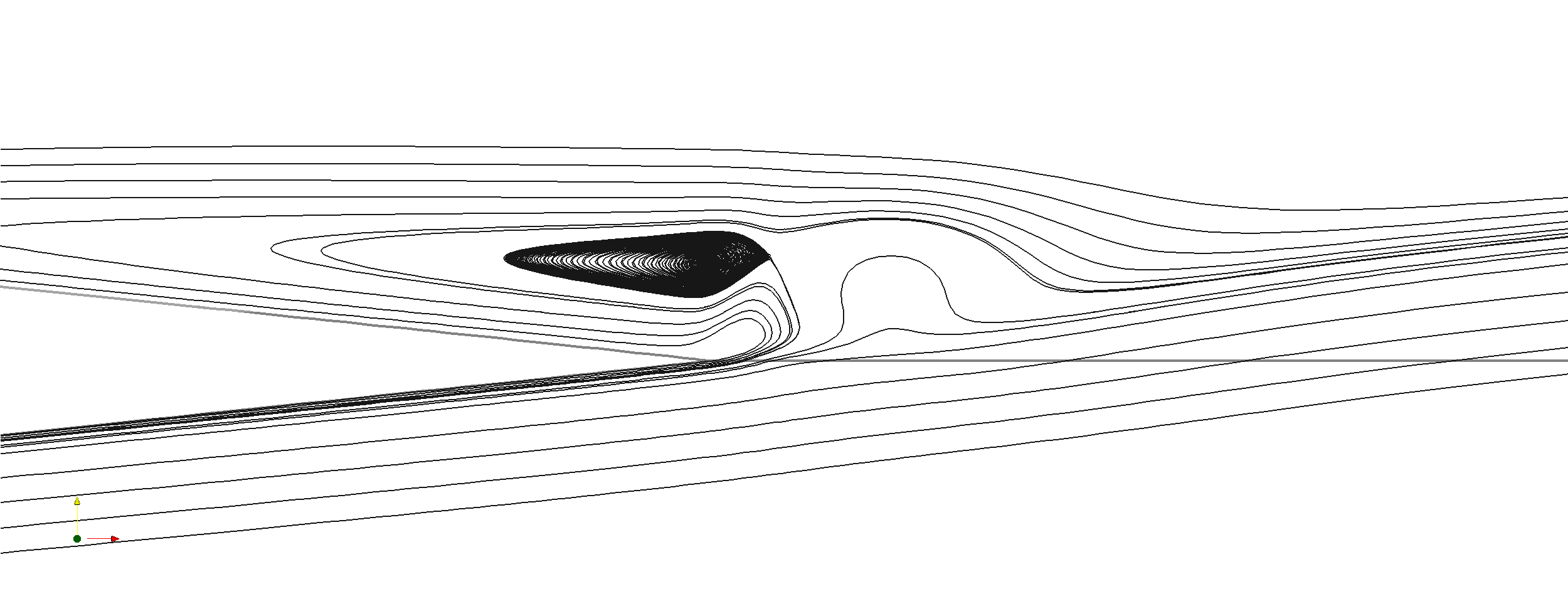}
\caption{Airfoil NACA 64A010, $Re = 2.0\times 10^6$,  $M=0.75$, angle of attack $\alpha = 6.2^\circ$. Streamlines showing the detail of the secondary flow around the trailing edge, obtained from the GKS solution.  }
\label{fig:64a010gksd}
\end{figure}

The serie-6 airfoil NACA 64A010 has been investigated in transonic flow by Johnson  \cite{johnson1981transonic}.
The case  $Re = 2.0\times 10^6$,  $M=0.75$, angle of attack $\alpha = 6.2^\circ$ has been considered here and investigated with the Navier-Stokes and GKS schemes. 
This flow case include a large separation with the shock wave located at about $30\%$ of chord. 
Neither the Navier-Stokes nor the GKS manage to capture the shock position accurately (figure \ref{fig:naca64a010}). 
However, this case provides the evidence of a different prediction of vortical / turbulent flow. The re-circulation area by the trailing 
edge is predicted to have two different patterns by the Navier-Stokes (figure \ref{fig:64a010ns}) and GKS (figure \ref{fig:64a010gks}), especially around the trailing edge (figure \ref{fig:64a010gksd}). 
Surprisingly, Johnson envisages re-circulation pattern similar to the one predicted by the GKS scheme (\cite{johnson1981transonic}). 
The computational meshes are C-type with $576 \times 128$ cells, and a resolution of the boundary layer of $y_1^+ < 1$ in wall units.

\section{Conclusions}
\label{sec:conclusions}


The GKS investigated in this study (\cite{xu2001gas}) provides two main advantages with respect to ordinary Navier-Stokes schemes: higher-order fluxes and simultaneous treatment of transport and collisions.  
In the numerical experiments described, the GKS has in fact outperformed a Navier-Stokes scheme  in a number of flow cases, characterized by interaction between shock and  turbulent boundary layer. Interestingly, the resolution of secondary flows suggests that the GKS treats turbulence in a way similar to higher-order turbulence models (e.g. \cite{wallin2000explicit}), which are based on assumptions on the type of turbulence. 
 

The turbulent GKS seem to be a good candidate to investigate turbulent flow and the more so  in the rarefied / transition regime, such as for instance flow cases related to hypersonic flight, where ordinary schemes still fail to provide accurate and reliable results   (\cite{rumsey2009compressibility,roy2006review}).
Not only are GKS much better suited than Navier-Stokes schemes to handle flows with not negligible $Kn$, but they would also provide advantages in turbulence modelling. 
It is worth reminding that virtually all turbulence theories have been developed under the assumption that the turbulent timescales are much bigger than molecular ones. As a matter of fact, in  a flow characterized by $M=10$ and $Kn=0.001$ the two timescales might be comparable, i.e. $\tau_t \simeq \tau$.  
This means that ordinary Navier-Stokes schemes not only separate transport and collisions but they also miss the interactions between molecular and turbulent dynamics.

Even in the continuum regime the properties of GKS schemes are much less known than those of Navier-Stokes schemes:
many aspects of GKS  still need to be clarified and future activities might include the sensitivity of results to the type of turbulence model, reconstruction order and truncation order in the Chapman-Enskog expansion.
\section*{Acknowledgement}
The author is grateful to K. Xu for the information and the support provided.  

\bibliographystyle{plain}   

\bibliography{bibliobgk}

\begin{thebibliography}{10}

\bibitem{bhatnagar1954model}
P.L. Bhatnagar, E.P. Gross, and M.~Krook.
\newblock A model for collision processes in gases. i. small amplitude
  processes in charged and neutral one-component systems.
\newblock {\em Physical review}, 94(3):511, 1954.

\bibitem{cercignani1988boltzmann}
C.~Cercignani.
\newblock {\em The Boltzmann equation and its applications}, volume~67.
\newblock Springer, 1988.

\bibitem{chen2003extended}
H.~Chen, S.~Kandasamy, S.~Orszag, R.~Shock, S.~Succi, and V.~Yakhot.
\newblock Extended boltzmann kinetic equation for turbulent flows.
\newblock {\em Science}, 301(5633):633--636, 2003.

\bibitem{chen2004expanded}
H.~Chen, S.A. Orszag, I.~Staroselsky, and S.~Succi.
\newblock Expanded analogy between boltzmann kinetic theory of fluids and
  turbulence.
\newblock {\em Journal of Fluid Mechanics}, 519(1):301--314, 2004.

\bibitem{chou1997kinetic}
S.Y. Chou and D.~Baganoff.
\newblock Kinetic flux--vector splitting for the navier--stokes equations.
\newblock {\em Journal of Computational Physics}, 130(2):217--230, 1997.

\bibitem{cook1979aerofoil}
PH~Cook, MA~McDonald, and MCP Firman.
\newblock Aerofoil rae 2822--pressure distributions, and boundary layer andwake
  measurements. experimental data base for computer program assessment.
\newblock {\em AGARD Advisory}, 1979.

\bibitem{godunov1959difference}
S.K. Godunov.
\newblock A difference method for numerical calculation of discontinuous
  solutions of the equations of hydrodynamics.
\newblock {\em Matematicheskii Sbornik}, 89(3):271--306, 1959.

\bibitem{harris1981two}
C.D. Harris.
\newblock Two-dimensional aerodynamic characteristics of the naca 0012 airfoil
  in the langley 8 foot transonic pressure tunnel.
\newblock {\em NASA Technical Memorandum}, 1981.

\bibitem{jameson1983solution}
A.~Jameson.
\newblock Solution of the euler equations for two dimensional transonic flow by
  a multigrid method.
\newblock {\em Applied Mathematics and Computation}, 13(3-4):327--356, 1983.

\bibitem{johnson1981transonic}
DA~Johnson, WD~Bachalo, and FK~Owen.
\newblock Transonic flow past a symmetrical airfoil at high angle of attack.
\newblock {\em NASA Technical Memorandum}, 1981.

\bibitem{mandal1994kinetic}
JC~Mandal and SM~Deshpande.
\newblock Kinetic flux vector splitting for euler equations.
\newblock {\em Computers \& fluids}, 23(2):447--478, 1994.

\bibitem{may2007improved}
G.~May, B.~Srinivasan, and A.~Jameson.
\newblock An improved gas-kinetic bgk finite-volume method for
  three-dimensional transonic flow.
\newblock {\em Journal of Computational Physics}, 220(2):856--878, 2007.

\bibitem{pope2000turbulent}
S.B. Pope.
\newblock {\em Turbulent flows}.
\newblock Cambridge Univ Pr, 2000.

\bibitem{roy2006review}
C.J. Roy and F.G. Blottner.
\newblock Review and assessment of turbulence models for hypersonic flows:
  2d/axisymmetric cases.
\newblock {\em AIAA Paper}, 713:2006, 2006.

\bibitem{rumsey2009compressibility}
CL~Rumsey.
\newblock Compressibility considerations for kappa-omega turbulence models in
  hypersonic boundary layer applications.
\newblock {\em NASA Technical Memorandum}, 2009.

\bibitem{tang2011progress}
L.~Tang.
\newblock Progress in gas-kinetic upwind schemes for the solution of
  euler/navier-stokes equations i. overview.
\newblock {\em Computers \& Fluids}, 2011.

\bibitem{wallin2000explicit}
S.~Wallin and A.V. Johansson.
\newblock An explicit algebraic reynolds stress model for incompressible and
  compressible turbulent flows.
\newblock {\em Journal of Fluid Mechanics}, 403:89--132, 2000.

\bibitem{wilcox2006}
D.~C. Wilcox.
\newblock {\em Turbulence Modeling for CFD, 3rd edition}.
\newblock DCW Industries, Inc., La Canada CA, 2006.

\bibitem{xu2001gas}
K.~Xu.
\newblock A gas-kinetic bgk scheme for the navier--stokes equations and its
  connection with artificial dissipation and godunov method.
\newblock {\em Journal of Computational Physics}, 171(1):289--335, 2001.

\bibitem{xu2005multidimensional}
K.~Xu, M.~Mao, and L.~Tang.
\newblock A multidimensional gas-kinetic bgk scheme for hypersonic viscous
  flow.
\newblock {\em Journal of Computational Physics}, 203(2):405--421, 2005.

\bibitem{xu1994numerical}
K.~Xu and K.H. Prendergast.
\newblock Numerical navier-stokes solutions from gas kinetic theory.
\newblock {\em Journal of Computational Physics}, 114(1):9--17, 1994.

\bibitem{yoon1988lower}
S.~Yoon and A.~Jameson.
\newblock Lower-upper symmetric-gauss-seidel method for the euler and
  navier-stokes equations.
\newblock {\em AIAA journal}, 26(9):1025--1026, 1988.

\end{thebibliography}

\end{document}